\documentclass{article}
\usepackage[utf8]{inputenc}
\usepackage[a4paper, total={6in, 9in}]{geometry}
\usepackage{amsmath}
\usepackage{amsfonts}
\usepackage{cfr-lm}
\usepackage{yfonts}
\usepackage{bm}
\usepackage{graphicx}
\usepackage{float}
\usepackage{caption}
\usepackage{subcaption}
\graphicspath{ {./images/} }

\begin{document}

\begin{titlepage}
\centering
\begin{figure}[t]
    \includegraphics[width=9cm]{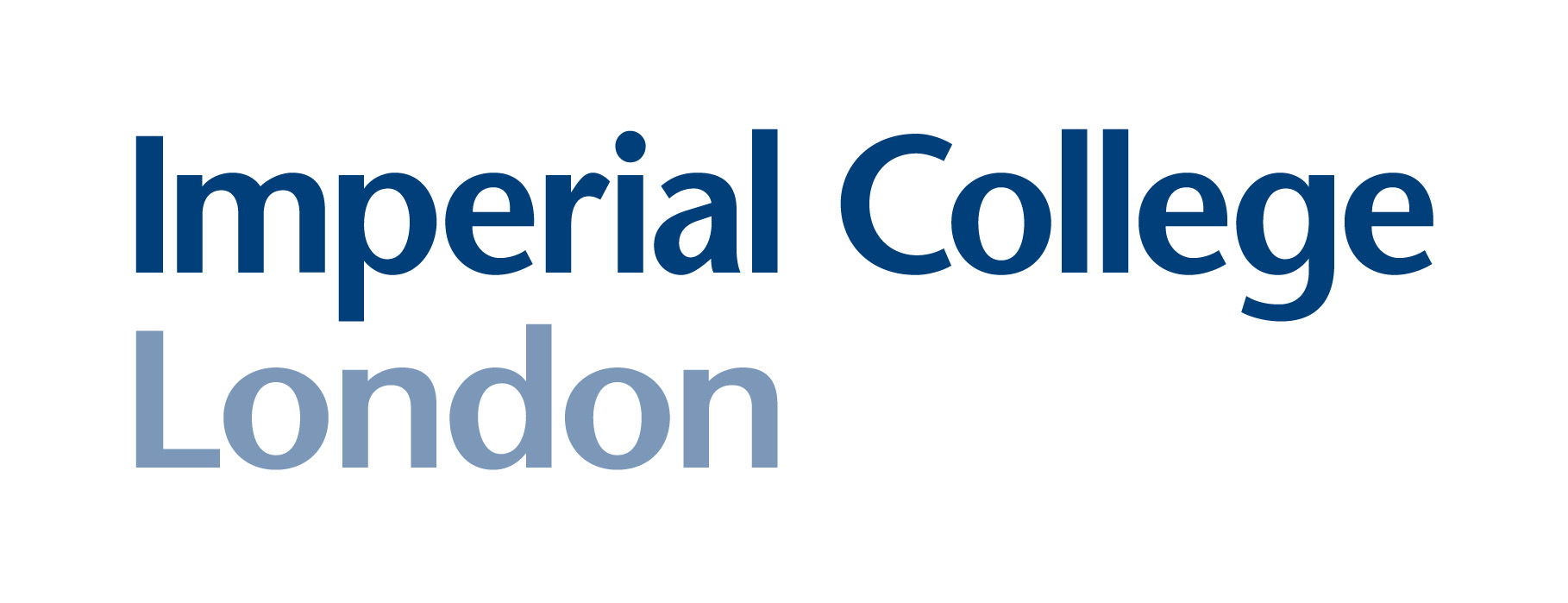}
\end{figure}
\vspace{1.5cm}
{\LARGE Imperial College London\\}
\vspace{0.2cm}
{\LARGE Department of Physics}\\
\vspace{1.5cm}
\noindent\rule{15cm}{0.4pt}\\
\vspace{0.3cm}
{\Huge\bfseries Quantum Cosmology of the Nothing}
\vspace{0.3cm}
\noindent\rule{15cm}{0.4pt}\\
\vspace{1cm}
{\Large Author: Alisha Marriott-Best\\}
\vspace{0.2cm}
{\Large Supervisor: Professor Jo\~{a}o Magueijo}
\end{titlepage}

\section*{\Large Acknowledgements}
First I would like to thank Professor Jo\~{a}o Magueijo, for his patience and guidance throughout this project. I'd also like to thank my family, and friends who have offered unwavering support to me and have helped me get to where I am today. Finally, I'd like to recognise the kind strangers of StackExchange and other internet fora who offered a great amount of assistance with Mathematica, R, and \LaTeX.
\newpage
\section*{Abstract}
Quantum cosmology uses a wave function to model the universe, but finding solutions for this poses a problem as it is difficult to define the boundary conditions or identify the correct path for a path integral. We begin the discussion by going over various proposals and look at how bubble universe nucleation can be used as an analogy for the tunneling wave function. We review how the Hartle-Hawking wave function and tunneling wave functions are equivalent. This leads into how the transition between the decelerated and accelerated expansion is formulated as a bounce in connection space. This is done in a toy model Universe that contains only radiation and a cosmological constant $\Lambda$. The wave function is a superposition on an incident wave, a reflected wave, and an evanescent wave; when it is constructed from wave packets. Using the toy model, we introduce the new concept of the universe tunneling to a different classical region during this bounce in connection space. This concept is explored by deriving the wave function of the evanescent wave $\psi_{ev}$ in order to calculate $|\psi|^2$ to give an indication of the probability of the universe tunneling to another classical region. 

\newpage
\tableofcontents
\newpage

\section{Introduction}\label{intro}
\numberwithin{equation}{section}
\pagenumbering{arabic}
Quantum cosmology looks at the universe as a wave function $\psi$ rather than a space-time. Looking at the universe from a quantum perspective first originated from DeWitt \cite{WDW}; the wave function can be obtained by solving the Wheeler-DeWitt equation \cite{WDW} with the correct boundary conditions. Alternatively it can be obtained by solving path integrals with the right paths.
As quantum cosmology developed it was found that a small closed universe can spontaneously appear from nothing \cite{vnothing, VBC}. This concept of nothing being no matter and no space-time.\\ \newline
We can define the wave function of the universe with components of all 3 metrics $h_{ij}(x)$ and matter field configurations $\phi(x)$, as it is defined on superspace,\\
\begin{equation}
    \psi[h_{ij}(x),\phi(x)]
\end{equation}\\
This obeys the Wheeler-DeWitt equation:
\begin{equation}\label{wdw}
    \mathcal{H}\psi[h_{ij}(x),\phi(x)] = 0
\end{equation}
Equation (\ref{wdw}) has an infinite number of solutions like most other differentiable equations. For it to be possible to find a solution to this, we need to specify boundary conditions in the superspace. Usually these boundary conditions are determined by the external system but for quantum cosmology we don't have anything external to the universe.\\
There are many proposals for what boundary conditions should be used for (\ref{wdw}), or what path integrals to use.\\
\newline
Hartle and Hawking \cite{H-H} proposed a Euclidean path integral over compact 4-geometries bound by the $h_{ij}(x)$ and $\phi(x)$. 
\begin{equation}\label{harthawk}
    \psi = \int^{(h, \phi)} [dg][d\phi]\exp[-S_E(g, \phi)]
\end{equation}
In quantum field theory a euclidean rotation of the time axis t $\rightarrow -i\tau$ is used as it improves the path integral convergence; in quantum gravity, the opposite happens. The path integral  by Hartle and Hawking is badly divergent due to $S_E$ being unbounded from below, and analytic continuation can only partly help solve this issue. This calls into question how meaningful this definition of the integral is.\\
\newline
Vilenkin \cite{VBC} proposed an integral over Lorentzian histories bound between a vanishing 3-geometry {\zeroslash} and $(h, \phi)$.
\begin{equation}\label{vilwavint}
    \psi = \int^{(h, \phi)}_\zeroslash [dg][d\phi]\exp[iS]
\end{equation}
Linde \cite{LBC} proposed taking t $\rightarrow +i\tau$ instead as this makes the path integral converge for the scale factor. That is all that is required for mini superspace models. However, for models with in-homogeneity and matter degrees of freedom, it diverges. This could be fixed with additional contour rotations but none have been proposed.\\
\newline
Halliwell and Hartle \cite{Hal-H} proposed a path integral over complex metrics, which aren't specifically Euclidean or Lorentzian. And although this path integral encompasses all the previous proposals and opens the door to newer ones, the space of complex metrics is very large and there is not an obvious contour integration to choose.\\
\newline
Vilenkin also formulated the tunneling-boundary condition \cite{VTBC1, VTBC2}, which is a boundary condition in superspace. This condition requires $\psi$ to only have outgoing waves at the superspace boundary. The weakness here being that the "outgoing waves" and the boundary are not rigorously defined. The Vilenkin path integral (\ref{vilwavint}) is the path integral version of this condition.\\
\newline
It's been shown that the Chern-Simons state is equivalent to the Hartle-Hawking and Vilenkin wave functions in mini-superspace \cite{JCS} which has implications on the theory of the "universe being created from nothing". By generalising the universe as being dominated by radiation and fluids (with a generic equation of state), the correct classical limit can be found \cite{J2110}. This is because the radiation and fluids have a different constant which means the chosen time is also different. The classical limit requires variable quantum time and that peaked wave-packets can be found. Additionally by examining the group speed and finding the equations of motion of the peak of the suitable wave functions \cite{J2110}, the semiclassical limit is still obtained in cross-over regions. \\
\newline
$b$ is the inverse comoving Hubble parameter; opposed to the expansion factor $a$ ($b = \dot{a}/N$ on-shell for lapse function N). It has gone from a decreasing function to an increasing function of time \cite{btranref}, associated with $\Lambda$ or a more general form of dark energy taking over causing the expansion of the universe. This turning point between the decelerated expansion of radiation and the accelerated expansion of dark energy is a bounce in the connection representation at late times- known as $b$-bounce. The bounce occurs at $b = b_0$. It is important to understand that the $b$-bounce is not the same as the possible bounce in the metric representation occurring at the Planck epoch. The wave function is a superposition of 3 coherent wave packets; an incident wave packet (-, radiation epoch); a reflected wave packet (+, $\Lambda$ epoch); and an evanescent wave packet, which exists in the classically forbidden region around the $b$-bounce. When the incident and reflected wave packets (both being in semiclassical states) interfere, they produce a "ringing" in the wave function and probability density function, but not the actual probability \cite{gielen}. This demonstrates how two semiclassical states don't produce a semiclassical state when they superpose. If you compute $|\psi|^2$ close to the bounce you will see the ringing effect. You can construct something similar using the wave packets locked to $T_\phi$, showing one clock changing to another.\\
\newline
Given the concept of the universe spontaneously nucleating, one would ask the question as to whether the universe could spontaneously de-nucleate and how probable that would be. Although the answer to these questions lies outside the realm of this paper, we do for the first time, examine how the universe could tunnel from one classical region to another. By considering the evanescent wave packet in the toy model of radiation and $\Lambda$ we calculate the probability density $|\psi|^2$ at $-b_0$. This will indicate the probability that the universe could have moved from the region where it is expanding; transitioning from deceleration in radiation to an acceleration in dark energy. Instead moving to a region where it is contracting; transitioning from decelerating dark energy to accelerating radiation.

\section{Bubble Nucleation}
This section of the paper is about bubble nucleation and how Vilenkin \cite{Vilenkinrev} formulates the outgoing-wave boundary condition for a nucleating bubble. The process takes place in a false vacuum which can be thought of as the nucleation of universes. Here the boundary condition is formulated using a spherical minisuperspace model. In order to do this a number of assumptions need to be made:
\begin{itemize}
    \item At nucleation the bubble radius R $\gg$ the bubble wall thickness. This means the bubble can be approximated to a thin sheet.
    \item Use semiclassical approximation - assuming that the tunneling action is large. (Which is always true for a bubble provided it is weakly interacting).
    \item Ignore the gravitational effects of the false vacuum and assume Minkowski spacetime.
\end{itemize}
The second assumption means that we can say the bubble is nearly spherical and can be described by a minisuperspace model with 1 degree of freedom, and bubble radius R.
In this model the worldsheet of the bubble wall is only described by R(t) and the energy of the system does not change due to bubble nucleation.\\
The world sheet metric is derived as
\begin{equation}
    ds^2 = (1-\dot{R}^2)dt^2 - R^2(t)d\Omega^2
\end{equation}
and the solution for R(t),
\begin{equation}
    R(t) = (R^2_0 + t^2)^{1/2}
\end{equation}
These can be re-written using a new time coordinate and recognising the metric is of (2+1) - dimensional de Sitter space:
\begin{equation}
    \tau = R_0 \sinh^{-1}\left(\frac{t}{R_0}\right)
\end{equation}
\begin{equation}
    ds^2 = d\tau^2 - R^2(\tau)d\Omega^2
\end{equation}
\begin{equation}
    R(\tau) = R_0\cosh{\left(\frac{\tau}{R_0}\right)}
\end{equation}
A 2-d being living in the walls of the bubble would be able to determine that they live in an expanding inflationary universe; it would also be possible for them to figure out their universe spontaneously nucleated at $\tau = 0$, so the metric and solution for R(t) that we have is for $\tau > 0$.
The discussion continues onto how the 2-d beings would describe the nucleation.
\begin{equation}
    H\psi = 0
\end{equation}
\begin{equation}\label{nuc bubble}
    [-\partial_R^2 + U(R)]\psi = 0
\end{equation}
The 2-d being would find a tunneling probability \cite{bub1, bub2, bub3} from:
\begin{equation}
    \bigg|\frac{\psi(R_0)}{\psi(0)}\bigg|^2 \sim \exp{\left( -2\int_0^{R_0}|p(R)|dR\right)} = \exp{\left(\frac{-\pi^2\sigma R_0^3}{2}\right)}
\end{equation}
This leads to some questions:
\begin{itemize}
    \item What does it mean to have a probability when there is only one bubble?
    \item Other bubbles would be unobservable - how could this be tested by observation?
\end{itemize}
For an observer of the bubble the probability would be well defined, but we don't have that for the universe. Nevertheless, it's possible for the worldsheet observers to derive useful information from their $\psi$. In the nothingness where multiple bubbles can nucleate, an observer is most likely to find themselves in a bubble with the highest nucleation probability. The nucleating bubbles don't have to be spherical and it would be possible to calculate the amplitude of which shape a bubble would have. The perturbative superspace approximation has solved this problem including all degrees of freedom of the bubble and treats all motions except for radial ones as small perturbations \cite{qsnb, qfb}. The perturbations can be seen as excitations of scalar field $\Phi$ that exists on the world sheet and has tachyonic mass $m^2 = -3R_0^2$.\\
\newline
This field has multiple modes, the mode expansion shows that it has 4 "$0$ modes", representing the time and space translations of the bubble. Other modes represent the deviations of the spherical shape. The field $\Phi$ nucleates in a de-Sitter invariant quantum state  which is the same as the cosmological case \cite{halhawk, utwfu}. It is possible to test this prediction by the external and worldsheet observers. Despite the lack of dispute about this result, it is not clear what form the boundary conditions take to select this wave function. The outgoing-wave boundary condition should be satisfied by the non-perturbative, radial part of $\psi$; the rest of $\psi$ should then be fixed depending on the condition $|\psi|<\infty$ \cite{qsnb, qfb}. It is emphasised in \cite{SasTanYamYok, SasTanYam} that the boundary condition needs to reflect that the bubble nucleates from a vacuum and not an excited state. But the form of boundary condition suggested in \cite{SasTanYamYok, SasTanYam} doesn't work for a thin wall bubble. There is a possibility for $\psi$ to be completely fixed provided that it respects the Lorentz invariance of the vacuum \cite{Vilenkinrev}.\\
\newline
The analysis of the bubble universe has not been expanded past the perturbative superspace, as it has not been solved. If the bubble worldsheet is represented by the parametric form $x^\mu(\xi^a)$ where $a = 0$, $1$, $2$. To find the amplitude of the bubble as an external observer in the configuration $x^0 = T$ the following path integral needs to be evaluated:
\begin{equation}
    \psi = \int [dx^\mu]e^{iS}
\end{equation}
This is a very difficult integral to calculate or make well defined.
\newline
It is to be expected that at small length scales there would be large quantum fluctuations; if large deformations are allowed then the bubble wall would be able to cross itself and daughter bubbles would be created. At very small scales it is possible that the bubble may have a fractal structure, with a dense layer of the daughter bubbles around it \cite{Vilenkinrev}. The 2-dimensional observers may be able to determine that they are not on a 2-d surface by looking at solutions to the (3+1)-d field equations.

\section{de Sitter Minisuperspace}
This section discusses a similar approach to the Robertson-Walker universe $\Lambda > 0$ in Vilenkin \cite{Vilenkinrev}.
Using the minisuperspace model,
\begin{equation}\label{minsupact}
    S = \int d^4x\sqrt{-g}\left(\frac{R}{16\pi G}-\rho_v\right)
\end{equation}
gives the equation of motion for (N=$1$):
\begin{equation}\label{eqmotN}
    \dot{a}^2 + 1 -\Lambda a^2 = 0
\end{equation}
The solution is the de Sitter space:
\begin{equation}\label{desitsol}
    a(t) = H^{-1}\cosh{(Ht)}
\end{equation}
where $H = \Lambda^{1/2}$.\\
This model is quantised by replacing $p_a \rightarrow -i\partial/\partial a$ and implementing the Wheeler-DeWitt equation again,
\begin{equation}\label{qmoddesit}
    \left[ \frac{d^2}{da^2} + \frac{\gamma}{a}\frac{d}{da} - U(a)\right]\psi(a) = 0
\end{equation}
where
\begin{equation}
    U(a) = a^2(1-\Lambda a^2)
\end{equation}
We see similarities between this and the nucleating bubble (\ref{nuc bubble}). The $\gamma$ dependent term has no affect on the wave function in the semiclassical regime. Omitting this term reduces this equation to the same form as the 1-d Schr\"{o}dinger equation for a particle that has zero energy and moves in the potential $U(a)$.
The momentum found for the action was
\begin{equation}\label{momac}
    p_a = -a\dot{a}/N
\end{equation}
Vilenkin shows that for the classically allowed region is $a \geq H^{-1}$, the WKB solutions are the solutions to (\ref{qmoddesit}). 
For the region $a \gg H^{-1}$ the solutions are
\begin{equation}\label{momaccH}
    \hat{p}_a\psi_\pm(a) \approx \pm p(a)\psi_\pm(a)
\end{equation}
where $p(a) = [-U(a)]^{\frac{1}{2}}$.\\
Equation (\ref{momac}) and (\ref{momaccH}) show that $\psi_-(a)$ and $\psi_+ (a)$ describe a universe that expands and contracts.
In the tunneling picture we assume that the universe was originally very small and expanded to its very large size. That means that the wave function component that describes the universe contracting from the infinitely large size should absent:
\begin{equation}
    \psi(a>H^{-1}) = \psi_-(a)
\end{equation}
The WKB connection formula was used to find the under-barrier wave function,
\begin{equation}\label{undbarwav}
    \psi(a<H^{-1}) = \tilde{\psi}_+(a) - \frac{i}{2}\tilde{\psi}_-(a)
\end{equation}
Away from the classical turning point $a=H^{-1}$, the first term of equation (\ref{undbarwav}) dominates and so the nucleation probability is approximated as \cite{VBC, LBC},
\begin{equation}
    \bigg|\frac{\psi(H^{-1})}{\psi(0)}\bigg|^2 \sim \exp{\left( -2\int_0^{H^{-1}}|p(a')|da'\right)} = \exp{\left(-\frac{3}{8G^2\rho_v}\right)}
\end{equation}
\\
The time coordinate can be thought of as an arbitrary label in GR, it is only convention to choose time growing/decreasing towards the future. Once the convention is set, the tunneling wave function for this universe model is uniquely defined.\\
Here we define "future" by the growth of entropy or by the expansion of the universe.\\
\newline
There is a "generic" boundary condition suggested by Strominger \cite{stro}. He argued that the boundary condition should be imposed on $\psi$ at small $a$ rather than large $a$ because at nucleation the universe is governed by small-scale physics, which is similar to the tunneling approach.\\
\newline
Because the under-barrier wave function is generally a linear combination of $\tilde{\psi}_+(a)$ and $\tilde{\psi}_-(a)$, at the boundary condition of $a = 0$ you expect that the terms to be comparable.
However it can be seen that the positive term decreases exponentially with $a$, but the negative term grows exponentially making it dominate for all but infinitesimally small $a$.\\
The wave function for the classically allowed range using the WKB connection formula:
\begin{equation}\label{wav<H}
    \psi(a<H^{-1}) = \tilde{\psi}_-(a)
\end{equation}
\begin{equation}
    \psi(a>H^{-1}) = \psi_+(a) - \psi_-(a)
\end{equation}
This wave function is the same \cite{hawk} as if you applied the Hartle-Hawking prescription to this model. Due to the expanding and contracting components of (\ref{wav<H}) it feels natural to interpret it as a de Sitter universe (\ref{desitsol}) that is contracting and expanding. An alternate interpretation \cite{Don, rub} is to see $\psi_-(a)$ and $\psi_+$ as the time reverses of each other. Whilst they describe the same nucleating universe, they have different time coordinate directions. One issue with this interpretation is that problems arise when the two components interfere.\\
\newline
This case of applying the boundary conditions at small $a$ is unconvincing, even though it can be applied to the bubble nucleation we know that the correct boundary condition is the outgoing wave at large radii. Similar to the boundary conditions imposed at infinity for bound states in a hydrogen atom.

\section{Tunneling Wave Function by Analytic Continuation}\label{tunwavana}
This section discusses how the wave function is found from a "bound-state" universe in Vilenkin \cite{Vilenkinrev}.
To find the quantum mechanical wave function for the decay of a metastable state, the bound state wave function can be analytically continued. This same approach can be used in quantum cosmology. This example uses the same minisuperspace model (\ref{minsupact}), but with $\rho_v < 0$. In this case $\Lambda < 0$, which means that the equation of motion (\ref{eqmotN}) has no solutions. But Planck-size universes could still come out and then collapse as quantum fluctuations. It is then expected that the wave function would peak at very small scales and collapse at $a \rightarrow \infty$.\\
It is necessary to use exact solutions to (\ref{qmoddesit}), which you can get \cite{VQOU} from choosing the factor ordering parameter, $\gamma = -1$. With the boundary condition,
\begin{equation}\label{tunwavbc}
    \psi(a\rightarrow\infty)=0
\end{equation}
The solution is the \textbf{Airy function}
\begin{equation}
    \psi(a)=Ai(z)
\end{equation}
\begin{equation}
    \tilde z = (-2\Lambda)^{-2/3}(1-\Lambda a^2)
\end{equation}
Using the behaviour of the wave function and using the relation \cite{abram} it is concluded that the wave function for $\Lambda > 0$ is:
\begin{equation}
    \psi(a) = A\mathrm{i}(\tilde{z}) + iB\mathrm{i}(\tilde{z})
\end{equation}
which is the tunneling wave function \cite{VTBC2}. There is an important difference between the standard treatment of the decay of the metastable state and what has just been done. Usually the Schrodinger equation for a bound state,
\begin{equation}
    \mathcal{H}\psi = E\psi
\end{equation}
is solved using the boundary conditions $\psi \rightarrow 0$, at $x \rightarrow 0$ and $x \rightarrow -\infty$. The eigenvalues are then completely determined by the Hamiltonian. This means the resulting wave functions describe a probability that exponentially decreases with time inside the potential well. In the quantum cosmology model the eigenvalue of the Wheeler-DeWitt operator is fixed at $E=0$, the wave function is defined on a half-line $a>0$ and the boundary condition (\ref{tunwavbc}) is applied at $a \rightarrow \infty$. The wave function is time independent, and the steady probability flux at $a \rightarrow \infty$ is sustained by the flux coming in at $a=0$. The tunneling wave function is more like the Green's function with source at a=0.\\
\newline
For $\Lambda > 0$ choosing a generic boundary condition at $a = 0$ would make the wave function not be confined and increase without bound at $a \rightarrow \infty$.

\section{Tunneling Wave Function from a Path Integral}\label{tunwavpath}
This section looks at a more complicated minisuperspace model- the Robertson-Walker universe with a homogeneous scalar field. This is done in \cite{Vilenkinrev} to demonstrate the relation between the out-going wave and path-integral forms for the tunneling wave.\\
The path integral (\ref{vilwavint}) for this model is expressed as \cite{Teit, HalliLou}
\begin{equation}
    K(q_2, q_1) = \int_0^\infty dTk(q_2, q_1; T)
\end{equation}
\begin{equation}
    k(q_2, q_1; T) = \int_{q1}^{q2}[dq]\exp{\left(i\int_0^T \mathcal{L} dt\right)}
\end{equation}
This satisfies the Schr\"{0}dinger equation
\begin{equation}
    \left(i\frac{\partial}{\partial T} - \mathcal{H}_2\right)k(q_2, q_1; T) = 0
\end{equation}
with initial condition
\begin{equation}
    k(q_2, q_1; 0) = \delta(q_2,q_1)
\end{equation}
The equation for $K(q_2, q_1)$ follows from these
\begin{equation}
    \mathcal{H}_2K(q_2, q_1)=-i\delta(q_2,q_1)
\end{equation}
Here $\mathcal{H}$ is the Wheeler-DeWitt operator
\begin{equation}
    \mathcal{H}=\frac{1}{2}e^{-3\alpha}[\partial^2_\alpha - \partial^2_\phi - U(\alpha, \phi)]
\end{equation}
This has the "superpotential"
\begin{equation}
    U(\alpha, \phi) = e^{4\alpha}[1-e^{2\alpha}V(\phi)]
\end{equation} 
Ignoring the factor-ordering ambiguity, the subscript of 2 on $\mathcal{H}$ indicates that $\alpha$ and $\phi$ are taken as $\alpha_2$, $\phi_2$.\\
The operator $\mathcal{H}$ is just the KG operator for a relativistic particle in a (1+1) dimensional spacetime, with $\phi$ being the space coordinate and $\alpha$ being the time coordinate. The particle moves in an external potential $U(\alpha, \phi)$. Now considering the behaviour of K($q_2,q_1$) as $\alpha_2 \rightarrow \infty$ where $\alpha_1$ is fixed. As $\alpha \rightarrow -\infty$ the potential vanishes, and K($q_2,q_1$) is given by the superposition of plane waves $\exp[ik(\alpha_2\pm\phi_2)].$ This should only include waves where k > $0$.\\
As $\alpha_2 \rightarrow \infty$ we see that U$(\alpha,\phi)$ diverges and the WKB approximation becomes more and more accurate. The dependence of K($q_2,q_1$) on $q_2$ is found from the superposition of terms $e^{iS}$. S is a solution to the Hamilton-Jacobi equation
\begin{equation}
    \left(\frac{\partial S}{\partial \alpha}\right)^2 - \left(\frac{\partial S}{\partial \phi}\right)^2 + U(\alpha, \phi) = 0
\end{equation}
S$(\alpha, \phi)$ describes the congruence of classical paths with
\begin{equation}
    \frac{d\phi}{d\alpha} = - \frac{\partial S/\partial \phi}{\partial S/\partial\alpha}
\end{equation}
Since $V(\phi) > 0$, $U(\alpha, \phi) \approx -e^{6\alpha}V(\phi) < 0$, we see the trajectories of the particle are asymptotically timelike and correspond to an expanding or contracting (from infinity) universe with $p_\alpha = \frac{\partial S}{\partial \alpha} < 0$ and $\frac{\partial S}{\partial \alpha} > 0$ respectively. As for $V(\phi) < 0$, the trajectories are asymptotically spacelike and are not able to extend to timelike infinity $i_+$ or null infinity $\mathcal{I}_+$, so $K(q_2, q_1) \rightarrow 0$ when $q_2$ is at $i_+$ or $\mathcal{I}_+$.\\
This shows that $K(q_2, q_1)$ satisfies the boundary conditions at $\alpha \rightarrow \pm \infty$. The tunneling wave function is obtained by setting $\alpha_1 \rightarrow -\infty$ and integrating over all initial values of $\phi$
\begin{equation}
    \psi(\alpha, \phi) = \int_{-\infty}^\infty d\phi'K(\alpha, \phi|-\infty,\phi')
\end{equation}
Now the trajectories originate from $i_-$ the past timelike infinity, the behaviour of $\psi$ on the rest of the superspace boundary should be the same as K($q_2,q_1$). The probability flux enters at $i_-$ and exits as outgoing waves at $i_+$ and $\mathcal{I}_-$. This means that the path integral and outgoing waves of the tunneling wave function are equivalent.

\section{Beyond Minisuperspace}
The difficulties we have with finding the outgoing wave boundary condition are similar to the difficulties we have with defining positive-frequency modes in general curved spacetime. The minisuperspace model in Section \ref{tunwavpath} was defined based on the general properties of the potential $U(\alpha, \phi)$ (the fact it has unbounded growth as $\alpha \rightarrow +\infty$, and that it vanishes as $\alpha \rightarrow -\infty$.\\
It is possible to verify that the Wheeler-DeWitt equation has similar properties, by writing it as \cite{WDW}:
\begin{equation}\label{rewdw}
    (\nabla^2 - U)\psi = 0
\end{equation}
With the superspace Laplacian:
\begin{equation}\label{supspalap}
    \nabla^2 = \int d^3xN\bigg[ G_{ijkl}\frac{\delta}{\delta h_{ij}}\frac{\delta}{\delta h_{kl}} + \frac{1}{2}h^{-1/2}\frac{\delta^2}{\delta\phi^2}\bigg]
\end{equation}
N is the lapse function in 3+1 decomposition of spacetime and $h_{ij}$ is the 3-metric
The superspace metric is given by:
\begin{equation}
    G_{ijkl} = \frac{1}{2}h^{-1/2}(h_{ik}h_{jl} + h_{il}h_{jk} - h_{ij}h_{kl})
\end{equation}
And the superpotential is:
\begin{equation}\label{suppot}
    U = \int d^3xNh^{1/2}[-R^{(3)}+\frac{1}{2}h^{ij}\phi_{,i}\phi_{,j} +V(\phi)]
\end{equation}
In equation (\ref{rewdw}) we see that U becomes negligible as $\alpha \rightarrow -\infty$ \cite{Vilenkinrev}. Wald \cite{wald} has suggested (in a different context) that it may be possible to define outgoing modes analogous to the plane waves from Section \ref{tunwavpath}.\\
This next part illustrates this idea in a reduced superspace model including all degrees of freedom of the scalar field $\phi$, and 1 gravitational variable $\alpha$.
The scalar field is written in the form \cite{Vilenkinrev}:
\begin{equation}
    \phi(x) = (2\pi^2)^{1/2}\sum_n f_nQ_n(x)
\end{equation}
$Q_n(x)$ is the harmonics on a 3-sphere. By re-writing (\ref{supspalap}) as \cite{halhawk, utwfu},
\begin{equation}
    \nabla^2 = e^{-3\alpha}\left(\frac{\partial^2}{\partial\alpha^2}-\sum_n\frac{\partial^2}{\partial f_n^2} \right)
\end{equation}
We get the asymptotic plane wave solutions are
\begin{equation}
    \psi(\alpha,f_n) = \exp{(ik_\alpha \alpha+i\sum_nk_nf_n)}
\end{equation}
where
\begin{equation}
    k^2_\alpha - \sum_n k^2_n = 0 
\end{equation}
The boundary condition is obtained at $\alpha \rightarrow -\infty$, at this limit the tunneling wave function only contains terms with $k_\alpha > 0$.\\
In order to find the tunneling condition on the remainder of the superspace boundary we need to decide what metrics and matter fields should be included in the superspace. The way the Wheeler-DeWitt equation was used in (\ref{rewdw})-(\ref{suppot}) means that all configurations $\{h_{ij}(x), \phi(x)\}$ where $h^{-1/2}R^{(3)}$, $h^{-1/2}h^{ij}\phi_{,i}\phi_{,j}$, and $h^{-1/2}V(\phi)$ are integrable functions. The superpotential is then finite everywhere and will diverge towards the boundary.\\
\newline
When some dimensions of the universe become very large (e.g. $\alpha \rightarrow \infty$) the classical description of the degrees of freedom becomes more and more accurate. Let these classical variables be denoted by $c_i$; the remaining variables denoted by $q_j$; so that the wave function can be written as
\begin{equation}
    \psi(c, q) = \sum_N e^{iS_N(c)}\chi_N(c, q)
\end{equation}
\textbf{Note:} The superspace defined by $|U| < \infty$ contains a wide class of configurations
\begin{itemize}
    \item Metric and matter fields must be continuous (but don't have to be differentiable).
    \item Scalar fields that have discontinuous derivatives are acceptable configurations.
    \item Metrics with $\delta$-function curvature singularities are also acceptable configurations.
\end{itemize}
This fits with the path integral approach, where we know it is dominated by continuous paths that are not differentiable. It is possible to think of the superspace configurations as slices of these paths.
\\
If this sections definition of outgoing waves is possible, then the wave function from Section \ref{tunwavpath} defined by the path integral in Section \ref{intro} should satisfy the outgoing wave boundary condition. This path integral is advantageous as it is consistent even if the outgoing waves are undefined, and that the path integral is better suited to handling topology change \cite{Vilenkinrev}.
\newpage
\section{Topology Change}
The "creation of the universe from nothing" is an example of a topology changing event. The Wheeler-DeWitt equation (\ref{rewdw}) is based on canonical quantum gravity which assumes spacetime to be a manifold of topology $R \times \Sigma$, here R is a real line and $\Sigma$ is a closed 3-manifold of arbitrary but fixed topology.\\
It can be shown how superspace is divided into topological sectors by looking at lower dimensions. In the case of (1+1)-dimensions, 3-geometries get replaced by strings and the topological sectors are labeled by the occupation number of closed strings. In (2+1)-dimensions a point $g \in \mathcal{G}$ corresponds to many membranes and each surface is characterised by the number of handles \cite{Vilenkinrev}. In sections 3-5 the creation of the universe from nothing was described as a transition from a null topological sector to the sector with one universe with the topology $S_3$. The surface $\alpha = -\infty, |\phi| < \infty$ can be thought of as the boundary between these sectors.
If the topology change is a quantum tunneling event, then it can be represented by a smooth Euclidean manifold $\mathcal{M}$ that interpolates between the initial and final configurations. These configurations can be described using Morse theory \cite{Milnor}.\\
\newline
\textbf{Morse Function:} $f(x)$ is a Morse function if it has the following properties:
\begin{itemize}
    \item $0 < f(x) < 1$, where $f(x) = 0$ if $x \in \Sigma_1$ and $f(x) = 1$ if $x \in \Sigma_2$
    \item all critical points of $f$ are in the interior of $\mathcal{M}$ (not on the boundary) and are non-degenerate.
\end{itemize}
If the superspace configurations are obtained by taking slices of $\mathcal{M}$, then these slices can be obtained as surfaces of constant $f$. Different Morse functions will provide different slices. All the slices will have a smooth geometry, except those that pass through critical points. In \cite{VTBC2} Vilenkin adopts the point of view that the topology change is the probability flux being injected into the superspace through the null sector boundary. The flux then flows between each of the different sectors through the regular boundaries, and out of superspace through the singular boundary. However in \cite{Vilenkinrev}, Vilenkin no longer agrees with this picture, as topology change doesn't have to happen in between the configurations at each of the boundaries in the sectors. It involves the configurations that lie in the interiors of the sectors.\\
\newline
An example of lower dimensional topology change is the re-connection of intersecting strings. This process is crucial in the evolution of cosmic strings \cite{cosstrings}, at the classical level. But at the quantum level it instead represents the elementary interaction vertex in fundamental string theories. To summarise, topology changing transitions occur between points in the interior of different topological sectors as well as affecting the superspace boundaries. This implies that the Wheeler-DeWitt equation needs to be modified. In \cite{Vilenkinrev} Vilenkin proposed the addition of an operator $\tilde\delta$, which has matrix elements between the different superspace sectors.\\
\newline
Provided outgoing waves are defined along the previous section lines and the flux conservation condition is formulated then the wave function that is defined in 
\begin{equation}
    \mathcal{H}\psi_N(h) + \sum_{N'\neq N}\int[dh']\tilde\delta_{NN'}(h,h')\psi_{N'}(h')=0
\end{equation}
is hopefully equivalent to the path integral from Section 1.\\
It has been established that any Lorentzian metric that interpolates between two compact spacelike surfaces of different topology must be singular or contain closed timelike curves \cite{Geroch}. Since singularities can be mild \cite{Horowitz}, the corresponding spacetimes don't need to be excluded from the path integral. Including all finite action metrics is sufficient enough to allow for a Lorentzian topology change.

\section{Chern-Simons State Equivalence}
In \cite{JCS} the Chern-Simons state is shown to be equivalent to the  Hartle-Hawking equation (\ref{harthawk}) and tunneling wave functions. It is well established that the Chern-Simons state solves the full, non-perturbative Hamiltonian constraint in the self dual \cite{Jackiw, Freidel}. The Chern-Simons state is written as:
\begin{equation}\label{csstate}
    \psi(A) = \mathcal{N}\exp\left(-\frac{3}{2l^2_p\Lambda}Y_{CS}\right)
\end{equation}
where $\Lambda$ is the cosmological constant, $l^2_p = 8\pi G_N\hbar$, and the Chern-Simons functional $Y_{CS}$ is given by
\begin{equation}\label{csfunc}
    Y_{CS} = \int \mathcal{L}_{CS} = \int A^IdA^I + \frac{1}{3}\epsilon_{IJK}A^IA^JA^K.
\end{equation}
$A^I$ is the SU(2) Ashtekar self-dual connection.\\
\newline
There are some main criticisms against the Chern-Simons state \cite{witten2003note}:
\begin{itemize}
    \item It is non-normalisable.
    \item It has CPT violating properties.
    \item It has no gauge invariance for large gauge transformations.
\end{itemize}
These criticisms are based on the fact that the phase of the state is not purely complex; however in the minisuperspace approximation the state's phase is \textit{always} purely complex \cite{JCS}.\\
When the Einstein-Cartan action is reduced to minisuperspace we get a very simple Hamiltonian system \cite{varylambda, parityviol}.\\
The action:
\begin{equation}\label{csaction}
    S = 3\kappa V_c\int dt\left( 2a^2 \dot b + 2Na\left(b^2 + k -\frac{\Lambda}{3}a^2\right)\right)
\end{equation}
Where $\kappa = \frac{1}{16\pi G_N}$, $a$ is the expansion factor, on-shell $b \approx \dot a$ when there's no torsion, $k = 0,$ $\pm 1$ is curvature, and $V_c$ is the comoving volume of the region being studied.\\
The Hamiltonian equation is found, keeping the ordering from (\ref{csstate}), to be
\begin{equation}\label{hamcon}
    \hat{\mathcal{H}}\psi = \left(\frac{i\Lambda l_P^2}{9V_c}\frac{d}{db} +k+b^2\right)\psi = 0
\end{equation}
Which has the general solution:
\begin{equation}\label{hamcongs}
    \psi_{CS} = \mathcal{N}\exp \left[i\left(\frac{9V_c}{\Lambda l_P^2}\left(\frac{b^3}{3} +kb\right)+\phi_0\right) \right]
\end{equation}
Although there is ambiguity in $\phi_0$, it is set to $0$ as it will not affect the calculations.\\
This minisuperspace Hamiltonian (\ref{hamcon}) is the Wheeler-DeWitt equation (\ref{wdw}) in the complementary representation, which is implied from
\begin{equation}
    \left[\hat{b},\hat{a}^2\right] = \frac{il_P^2}{3V_c}.
\end{equation}
Equation (\ref{csaction}) is the Chern-Simons state reduced to minisuperspace; and equation (\ref{csfunc}) is the minisuperspace reduction of the Hamiltonian constraint. From the minisuperspace perspective the quantum cosmology base variable in the metric representation should be $a^2$ - this has radical implications.\\
Using the $a^2$ representation instead then the Hamiltonian constraint is:
\begin{equation}\label{a2hamcon}
    \left[ \frac{d^2}{da^2} - \frac{1}{a}\frac{d}{da} - U(a)\right]\psi = 0
\end{equation}
where
\begin{equation}
    U(a) = 4\left(\frac{3V_c}{l_P^2}\right)^2a^2\left(k-\frac{\Lambda}{3}a^2\right)
\end{equation}
since 
\begin{equation}
    \hat{b} = \frac{il^2_p}{3V_c}\frac{d}{d(a^2)}.
\end{equation}
Equation (\ref{a2hamcon}) is the Wheeler-DeWitt equation with a specific order. By setting $k=1$, $V_c = 2\pi^2$ and choosing $\alpha = -1$ like in \cite{Vilenkinrev} then it can be seen that there's an agreement. This is because the Einstein-Cartan action reduces to the Einstein-Hilbert space if there is no torsion. The Hartle-Hawking \cite{H-H, Vilenkinrev}, and the tunneling wave functions \cite{VTBC2,Vilenkinrev} are solutions of this equation; the solution that is obtained is dependent on the boundary conditions used.\\
The Hartle-Hawking and Vilenkin wave functions have to be related to the Chern-Simons state by a Fourier transform, which is inferred from (\ref{hamcon}), 
\begin{equation}\label{fourtrans}
    \psi_{a^2}(a^2) = \frac{3V_c}{l_P^2}\int \frac{db}{\sqrt{2\pi}}e^{-i\frac{3V_c}{l_P^2}a^2b}\psi_b(b)
\end{equation}
When the Wheeler-DeWitt equation is in the metric representation it is second order which allows for two linearly independent solutions (Hartle-Hawking and Vilenkin) but then in the $b$ representation it is only first order so the Chern-Simons wave function is unique. This means there's an issue with the Fourier transform; resolving this will show how the Chern-Simons state can be dual to the Hartle-Hawking and Vilenkin functions. As discussed in previous sections and in \cite{VTBC2, Vilenkinrev} the solutions are Airy functions. For the Vilenkin boundary conditions:
\begin{equation}
    \psi_v \propto A\mathrm{i}(-z) + iB\mathrm{i}(-z)
\end{equation}
and for the Hartle-Hawking boundary conditions:
\begin{equation}
    \psi_H \propto A\mathrm{i}(-z).
\end{equation}
Using the known results for Airy functions \cite{glinka, abram2}, in the integral representation,
\begin{equation}\label{airyint}
    \phi(z) = \frac{1}{2\pi}\int e^{i\left(\frac{t^3}{3}+zt\right)}dt
\end{equation}
Taking the Chern-Simons state (\ref{hamcongs}) and putting it into (\ref{fourtrans}) gives the integral (\ref{airyint}), with the following substitutions:
\begin{equation}
    z = - \left(\frac{9V_c}{\Lambda l_P^2}\right)^{2/3}\left(k-\frac{\Lambda a^2}{3}\right)
\end{equation}
and
\begin{equation}
    t = \left(\frac{9V_c}{\Lambda l_P^2}\right)^{1/3}b
\end{equation}
This means the Chern-Simons state is the Fourier dual of the Hartle-Hawking and Vilenkin wave functions \cite{JCS}.

\section{Dealing With Crossover Regions}\label{crossreg}
In the introduction we discussed how the wave function is constructed of 3 wave packets- incident, reflected, and evanescent. Again putting emphasis on the fact that at the bounce at the point of interference between the incident and reflected waves, is a bounce in connection space and not metric space.

\subsection{The Connection Space Picture}\label{bbounce}
Here we will briefly discuss how and why the connection picture is used, which is also used in \cite{J2110, gielen}. Any quantum cosmology description that excludes time struggles to connect with any of the classical descriptions. By using the connection representation we can recover classical results from semiclassical wave functions \cite{J2110}.\\
The connection representation uses the following description:
\begin{itemize}
    \item The minisuperspace connection variable $b$ is used instead of the expansion factor $a$ for the dependent variable.
    \item The physical time(s) $T_\alpha$ is used instead of the coordinate time $t$ for the independent variable.
\end{itemize} 
Recall from Section \ref{intro} that on-shell $b = \dot{a}/N$. When viewed quantum mechanically $b$ is a variable that is independent and complementary to the metric. Using $T_\alpha$ means that quantum mechanically we can have multiple times that, when described in the classical limit, are functions of $t$. This means there's only one time classically but multiple times quantum mechanically \cite{J2110}.\\
It can be shown that the classical trajectory for a single fluid system is:
\begin{equation}
    \dot{X}_\alpha = \dot{T}_\alpha.
\end{equation}
We need to assume the first Friedman equation:
\begin{equation}
    b^2 + k = \frac{m}{a^{1+3w}}
\end{equation}
and the two dynamical Hamiltonian equations:
\begin{equation}
    \dot{a}= \{a,H\}=Nb
\end{equation}
\begin{equation}
    \dot{b}= \{b,H\}=-\frac{1+3w}{2}Na\frac{m}{a^{3(1+w)}}
\end{equation}
Additionally for this description there are some things to note \cite{J2110},
\begin{itemize}
    \item For $b>0$ the universe is expanding, for $b<0$ the universe is contracting, and for $b=0$ the universe is static Close to $b=0$ will be the loitering models.
    \item For Phoenix or "bouncing" cosmologies \cite{Ekpyrotic, Phoenix}, $b$ will go through zero. It may be possible for tunneling to occur between branches that have different signs.
    \item If $w<-\frac{1}{3}$ then $b$ will increase in parameter time $t$ for a given matter content. In the $w>-\frac{1}{3}$ case $b$ will decrease. When $w=-\frac{1}{3}$ then $b$ will not change.
\end{itemize}
This last point demonstrates that the $b$-bounce is the transition between the decelerated and accelerated transition. At the end of inflation, there will be a "turn around" in the connection space which is the opposite of the $b$-bounce.
\subsection{Mono-chromatic Solutions}
It is shown in \cite{J2110} that the semiclassical limit can still be obtained when the wave function is sharply peaked.\\
Using the Hamiltonian
\begin{equation}
    H = Na\left(-(b^2 + k) + \frac{a^2}{\phi}+\frac{m}{a^2}\right)
\end{equation}
which spans across 2-dimensional space with
\begin{equation}
    \alpha = \left(\phi\equiv\frac{3}{\Lambda}, m\right)
\end{equation}
The equation
\begin{equation}
    \mathcal{H}_0=h_\alpha(b)a^2-\alpha = 0
\end{equation}
can be used to solve for spatial $\psi_s$, by setting $H=0$ with $\alpha = \phi$. Solving the resulting quadratic equation in $a^2$ gives the solution
\begin{equation}
    a^2 = \frac{g\pm\sqrt{g^2-4m/\phi}}{2/\phi}
\end{equation}
Since $a^2 \in\mathbb{R}$ but can be positive or negative then 
\begin{equation}
    g^2\geq g_0^2 = \frac{4m}{\phi} = \frac{4}{3}\Lambda m
\end{equation}
When $g^2 >> g_0^2$ we get two branches, the minus one is radiation dominated and the plus one is Lambda dominated. But at $g^2 \approx g_0^2$ there's a  transition between decreasing b and increasing b which is decelerated expansion to accelerated expansion.\\ $\psi_s$ is written as
\begin{equation}
    \psi_{s\pm}(b;\phi,m)=\mathcal{N} \exp\left[i\frac{3V_c}{l_P^2}\phi X_\pm(b;\phi,m)\right]
\end{equation}
The limits obtained for $g^2 \gg \frac{m}{\phi}$ of the $\pm$ branches are:
\begin{equation}
    \psi_{s+}(b;\phi, m) \approx \mathcal{N} \exp\left[i\frac{3V_c}{l_P^2}\phi X_\phi(b)\right]
\end{equation}
\begin{equation}
    \psi_{s-}(b;\phi, m) \approx \mathcal{N} \exp\left[i\frac{3V_c}{l_P^2}mX_r(b)\right]
\end{equation}
This shows that $\psi_s(b; \alpha)$ is a piecewise plane wave. It appears that to leading order that if $\mathcal{A}(\alpha)$ factorises, then so will everything else and they become independent of b - but this is not the case.\\
The monochromatic solutions can be superposed into wave-packets, these packets move with a group speed as well as a phase speed.
\subsection{The Radiation Clock in the Lambda Epoch}
This is an examination of what happens to the "minority" clock once the crossover has finished \cite{J2110}.
For any factorisable amplitude:
\begin{equation}
    \psi(b,T) = F_1(X_\phi-T_\phi)F_2(X_r+T_r)
\end{equation}
For this clock, the wave function produces classical equations of motion, the peak moves along
\begin{equation}
    \dot{b} = \frac{a}{\phi} - \frac{m}{a^3}
\end{equation}
and has the classical trajectory,
\begin{equation}
    \dot{X}_r \approx -\dot{T}_r
\end{equation}
equivalently $\dot{X}_\phi \approx \dot{T}_\phi$. So the peak of the factors describes the classical trajectory. In quantum mechanics the two time components remain independent. For the peaked states the peak maps out the trajectory of $b$ in $2$D space.

\subsection{The Lambda Clock in the Radiation Epoch} 
In the radiation epoch \cite{J2110}
\begin{equation}
    \psi(b, T; \alpha) = \mathcal{N}\exp\left[-\frac{i}{\textfrak{h}}\left(m(T_r-X_r)+\phi\left(T_\phi-\frac{m^2}{\phi^2}\int\frac{db}{g^3}\right)\right)\right]
\end{equation}
The factor in the wave function that is associated with $\Lambda$ now depends on m. This means the wave packets don't factorise separately into Lambda and radiation factors. The motion of the peak is still the correct classical limit
\begin{equation}
    \dot{b} = -\frac{m}{a^3}
\end{equation}
We can factorise the wave function as,
\begin{equation}
    \psi(b, T; \alpha) \approx \psi_1(T_r - X_r)\psi_2(b, T_\phi; m_0)
\end{equation}
which supports having a single classical time.

\subsection{During the Bounce}
We still see a classical trajectory even in the $b$-bounce. For the Lambda wave packet the peak moves along \cite{J2110}
\begin{equation}
    \dot{b} = \frac{\dot{T}_\phi}{a^4/\phi^2}\left(\frac{a^2}{\phi} - \frac{m}{a^2}\right) = \frac{a}{\phi}-\frac{m}{a^3}
\end{equation}
For the $m$ wave packet the peak moves along  \cite{J2110}
\begin{equation}
    \dot{b} = -\dot{T}_r\left(\frac{a^2}{\phi} - \frac{m}{a^2}\right)
\end{equation}
This shows the correct classical limit is always obtained provided the wave functions remain peaked. This is further investigated in \cite{gielen} to see whether it is a good approximation.

\section{Cosmological Bounce in b}
\subsection{Classical Theory}
Taking a model of 2 cosmological clocks, where we model the clocks as perfect fluids with the equation of state parameters $w = \frac{1}{3}$ for radiation and $w = -1$ for dark energy. We can characterise the fluids by their energy density $\rho$ or their conserved quantity $\rho a^{3(w+1)}$ in minisuperspace. It is convenient to use the conserved quantity as it is canonically conjugate to a clock variable.\\
For the equations of motion for $\chi_1$ and $\chi_2$ we need to introduce
\begin{equation}\label{eqmocons}
    \phi = \frac{3}{\Lambda};    T_\phi = -3\frac{\chi_2}{\phi^2}
\end{equation}
as it is convenient to replace $\Lambda$ and $\chi_2$.\\
The action for gravity with two fluids can be varied to get the Hamiltonian constraint
\begin{equation}
    -(b^2 +k)+\frac{m}{a^2}+\frac{\Lambda}{3}a^2 = 0
\end{equation}
Using (\ref{eqmocons}) we can get 2 solutions for the constraint in terms of $a^2$
\begin{equation}
    a^2_\pm = \frac{\phi}{2}\left(V(b) \pm\sqrt{V(b)^2-4m/\phi}\right)
\end{equation}
and, as suggested in \cite{J2110, Joao2104}, this can be re-written in terms of the conserved quantity $\phi$ 
\begin{equation}
    h_\pm(b)a^2_\pm - \phi := \frac{2a^2_\pm}{V(b) \pm \sqrt{V(b)^2-4m/\phi}} - \phi = 0
\end{equation}
The - sign solution corresponds to radiation domination $(\phi m >> a^4)$ and the + sign solution corresponds to $\Lambda$ domination, this is checked by seeing which solution survives in the limits $m \rightarrow0$ or $\Lambda \rightarrow 0 (\phi \rightarrow \infty)$.\\
The point at which radiation and dark energy have equal energy densities $\phi m = a^4$ is the bounce in b, at $\dot{b} = 0$. The bounce happens when the Universe is still expanding. This happens when $V(b) = 2\sqrt{m/\phi}$ or 
\begin{equation}\label{bsquare}
    b^2 = b^2_0 :=2\sqrt{\frac{m}{\phi}} - k.
\end{equation}
For the Hamiltonian
\begin{equation}
    \mathcal{H}_\pm = \frac{3V_c}{8\pi G}\left(a^2-\frac{\phi}{2}\left(V(b)\pm \sqrt{V(b)^2-4m/\phi}\right)\right)
\end{equation}
we get
\begin{equation}
    \frac{db}{dt} = 1; \frac{d(a)^2}{dt}=\phi b\left(1\pm\frac{V(b)}{\sqrt{V(b)^2 -4m/\phi}}\right)
\end{equation}
These dynamics correspond to a gauge where $b$ plays the role of time and expresses the solution for $a^2$ in a relational form $a^2(b)$.

\subsection{Quantum Theory}
We quantise minisuperspace from promoting the Poisson brackets
\begin{equation}
    \{b,a^2\} = \{m,\chi_1\}=\{\Lambda,\chi_2\}=\frac{8\pi G}{3V_c}
\end{equation}
to
\begin{equation}
    \left[\hat{b},\hat{a}^2\right]=\mathrm{i}\frac{l_P^2}{3V_c}
\end{equation}
The representation diagonalising $b$ is chosen such that
\begin{equation}
    \hat{a}^2 = -\mathrm{i}\frac{l_P^2}{3V_c}\frac{\partial}{\partial b}=:-\mathrm{i}\textfrak{h}\frac{\partial}{\partial b}
\end{equation}
where $\textfrak{h}$ is shorthand for the "effective Planck parameter"\cite{Barrow}. When the Hamiltonian constrain is implemented in the metric representation, all of the matter contents share the same gravity-fixed kinetic term. They have differing equations of state $w$, which is shown with different powers on $a$ in the effective potential \cite{Vilenkinrev}.\\
However in the connection representation, the matter contents share the same gravity-fixed effective potential $V(b) = b^2+k$. The different matter components appear as different kinetic terms. Using a single Hamiltonian constraint that is quadratic in $a^2$,
\begin{equation}
    \frac{a^4}{\phi}-V(b)a^2+m=0
\end{equation}
We can quantise a Hamiltonian constraint that is written as a two-branch condition
\begin{equation}
    \hat{a}^2-\frac{\phi}{2}\left(V(b)\pm \sqrt{V(b)^2-4m/\phi}\right)=0
\end{equation}
This means the two branches link with the mono-fluid prescriptions when radiation or dark energy dominates linking with \cite{J2110, Joao2104}. Ultimately we have 2 quantum theories: one resulting in a second order formulation; and one resulting in a 2-branch first order formulation.\\
From the second order perspective, to get solutions that match with the first order theory, $b$ would need to be placed entirely on the left or right depending on which branch is being analysed. Therefore, the ordering in each formulation can never match.

\subsection{Solutions of the Second Order Formulation}
\cite{gielen} provides solutions to solving the model in the first and second orders. In this model it is harder to solve in the second order. In \cite{gielen} the following ansatz is chosen
\begin{equation}
    \psi(b,m,\phi) = \exp\left(\frac{\mathrm{i}}{2\textfrak{h}}\phi\left(\frac{b^3}{3}+bk\right)\right)\chi(b,m,\phi)
\end{equation}
A general solution to the problem is found and it is written using tri-confluent Heun functions \cite{specfuncs}. They then normalise the Huen functions, and define the boundary conditions in terms of power series about $b=0$. The solutions work well for "no-bounce" boundary conditions at $b=0$ in the classically forbidden region. The definition of the Huen functions used diverge badly the larger b gets, and so it is not useful when studying the classically allowed region. The fact that the functions use power series is the reason they diverge so badly, if they are analysed numerically instead then the solutions decay at large b.\\
If $|\psi|^2$ is said to be the probability density, it falls off as $1/b^2$ at large b. This means that the highest probability would surround the bounce $b=b_0$. It is tempting to relate this property to the coincidence problem of cosmology, as it suggests an observer is most likely to find themselves not too far from the bounce. Exact solutions to the second order theory are found by using semiclassical solutions. The solutions have plane waves in the allowed region and a growing or decaying exponential in the forbidden region, which is expected. The general lowest-order WKB solution to the second order theory was found to be,
\begin{equation}
    \psi = c_+(m,\phi)e^{\frac{\mathrm{i}}{\textfrak{h}}\int^bdb'a^2_+}+c_-(m,\phi)e^{\frac{\mathrm{i}}{\textfrak{h}}\int^bdb'a^2_-}
\end{equation}

\subsection{Solution for First Order Formulation}\label{10.4}
\begin{figure}
    \centering
    \includegraphics{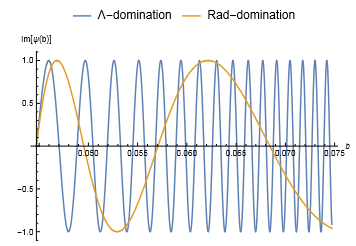}
    \caption{The imaginary part of $\psi_\pm$ for the two branches. The Lambda branch $\psi_{s+}$ is shown in blue, and the radiation branch $\psi_{s-}$ is shown in orange. It can be seen that the frequency increases with $b$ for the branches. Reference: \cite[Figure~3]{gielen}.}
    \label{fig:jbranches}
\end{figure}

\begin{figure}[ht]
    \centering
    \begin{subfigure}[b]{5cm}
        \centering
        \includegraphics[width=5cm]{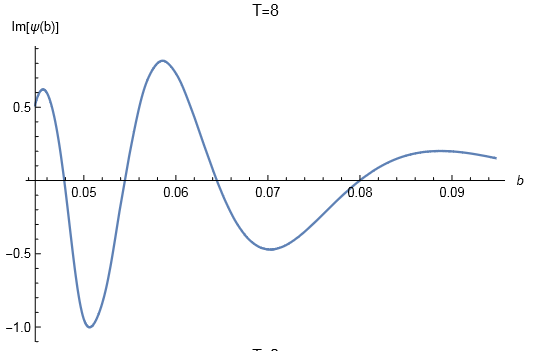}
        \label{fig:pos8}
    \end{subfigure}
    \begin{subfigure}[b]{5cm}
        \centering
        \includegraphics[width=5cm]{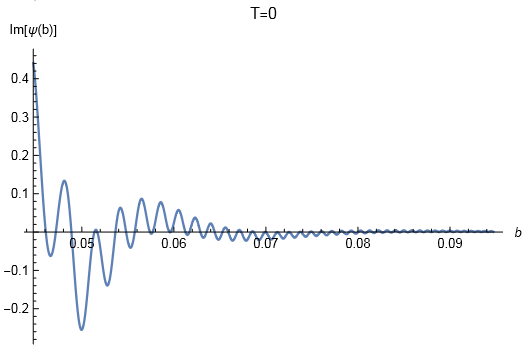}
    \label{fig:bounce}
    \end{subfigure}
    \begin{subfigure}[b]{5cm}
        \centering
        \includegraphics[width=5cm]{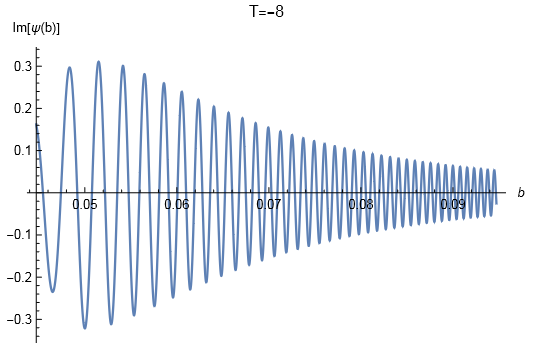}
        \label{fig:neg8}
    \end{subfigure}
    \caption{This shows the wave function in the classically allowed region $b\geq b_0$ either side of the bounce at $T=8$, $-8$. The correct part of the wave function is chosen from $\psi_s$ (+ or -) outside of the bounce. The $T=0$ graph (middle) shows that close to the bounce there is interference. Reference: \cite[Figure~7]{gielen}}
    \label{fig:claswav}
\end{figure}

\begin{figure}
    \centering
    \includegraphics{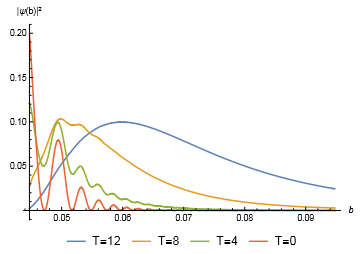}
    \caption{This plot shows the $|\psi|^2$ for the same situation as in Figure \ref{fig:claswav}. It is a symmetric function so $T_m < 0$ has been omitted from the plot. Reference: \cite[Figure~8]{gielen}}
    \label{fig:probdenclas}
\end{figure}

It is far easier to solve the first order formulation, \cite{gielen} provides a detailed solution. The general solution is given by solving the Wheeler-DeWitt with fixed values of $\alpha$, like in \cite{J2110}.
\begin{equation}
    \psi(b,T)=\int d\alpha \mathcal{A}(\alpha)\exp\left[-\frac{\mathrm{i}}{\textfrak{h}}\alpha \cdot T\right]\psi_s(b;\alpha)
\end{equation}
This is normalised in the conventional manner so that 
\begin{equation}
    |\psi_s|^2=\frac{1}{(2\pi\textfrak{h})^D}.
\end{equation}
This model has monochromatic solutions, where $\psi_s(b;\alpha)$ are the solutions of the two branches, given by
\begin{equation}
    \psi_{s\pm}(b;\phi,m)=\mathcal{N}\exp\left[\frac{\mathrm{i}}{\textfrak{h}}\phi X_\pm(b;\phi,m)\right]
\end{equation}
where
\begin{equation}
    X_\pm(b;\phi,m)=\int^b_{b_0} d\tilde b \frac{1}{2}\left(V(\tilde b) \pm \sqrt{V(\tilde b)^2-4m/\phi}\right)
\end{equation}
These solutions are plotted in Figure \ref{fig:jbranches}.\\
$b$ is defined by equation (\ref{bsquare}), and $b_0$ is taken to be the value of b at the bounce. These functions are plotted in Figure \ref{fig:jbranches}, using these limits and setting $\textfrak{h}=1$, $m=1$, $\phi = 10^6$, and $b_0 \approx 0.0447.$
A factorisable state is used to construct the wave packets,
\begin{equation}\label{factstate}
    \mathcal{A}(\alpha) = \prod_i \mathcal{A}_i(\alpha_i)=\prod_i\frac{1}{(2\pi\sigma^2_{\alpha i})^{1/4}}\exp\left[-\frac{(\alpha_i-\alpha_{i0})^2}{4\sigma^2_{\alpha i}}\right]
\end{equation} 
and then defining the spatial phases $P\pm$ from the wave function,
\begin{equation}\label{psis10}
    \psi_{s\pm}(b, \alpha) = \mathcal{N}\exp\left[\frac{\mathrm{i}}{\textfrak{h}}P_\pm(v,\alpha)\right]
\end{equation}
The wave function was found to be 
\begin{equation}\label{firstorwav}
    \psi_\pm(b,T)\approx \exp\left[\frac{\mathrm{i}}{\textfrak{h}}(P_\pm(b;\alpha_0)-\alpha_0\cdot T)\right]\prod_i\psi_{\pm i}(b, T_i)
\end{equation}
where
\begin{equation}
    \psi_{\pm i}(b,T_i) = \int\frac{d\alpha_i}{\sqrt{2\pi\textfrak{h}}}\mathcal{A}_i(\alpha_i)\exp\left[-\frac{\mathrm{i}}{\textfrak{h}}(\alpha_i-\alpha_{i0}\left(T_i-\frac{\partial P_\pm}{\partial \alpha_i}\bigg|_{\alpha_0}\right)\right]
\end{equation}
It was shown how in the classically allowed region, the motion of the peaks reproduces the classical equations of motion for the branches that was proven in \cite{J2110}. The motion is such that 
\begin{equation}
    X^{eff}_{\pm i}(b) = \frac{\partial P_\pm(b,\alpha)}{\partial\alpha_i}\bigg|_{\alpha_0}
\end{equation}
And so the packets are are of the form of Gaussians
\begin{equation}
    \psi_{\pm i}(b,T_i) = \frac{1}{(2\pi\sigma^2_{Ti})^{1/4}}\exp\left[-\frac{(X^{eff}_{\pm i}(b)-T_i)^2}{4\sigma^2_{Ti}}\right].
\end{equation}
As discussed in Section \ref{intro}, the wave function experiences ringing at the bounce. We demonstrated in Section \ref{crossreg}, with the full discussion in \cite{J2110} that the wave packet peaks follow the classical trajectory. But this does not apply at the bounce, due to the interference between the incident and reflected waves \cite{gielen}. This is evident in the middle graph of Figure \ref{fig:claswav} where it shows the ringing effect close to and at the bounce. There is also ringing in the probability function $|\psi|^2$ as shown in Figure \ref{fig:probdenclas}.
It is possible for this ringing to be observable \cite{gielen}. It needs to be checked whether the probability is associated with this probability density, and obtain the correct integration measure that ensures a unitary theory. The definition of the inner product shows how the ringing disappears in the semiclassical approximation, which is shown in the next section.
\newpage

\subsection{Inner Product and Probability Measure}
Due to the requirement of unitarity it is possible to infer the inner-product and probability \cite{gielen}. In mono-fluid situations there are 3 equivalent definitions as seen in \cite{J2110}:\\
\newline
\textbf{1) Mono-fluids}\\
A single fluid with equation of state parameter $w$ has a dynamical equation
\begin{equation}
    \left((b^2+k)^{\frac{2}{1+3w}}\frac{\partial}{\partial b}+\frac{\partial}{\partial T}\right)\psi =: \left(\frac{\partial}{\partial X}+\frac{\partial}{\partial T}\right)\psi = 0
\end{equation}
where $T$ depends on $w$ and $X = \int \frac{db}{(b^2+k)^{\frac{2}{1+3w}}}$, the inner product is inferred from this, and defined as
\begin{equation}
    \langle \psi_1|\psi_2\rangle = \int dX(\psi_1^*(b(X), T)\psi_2(b(X), T))
\end{equation}
It is assumed that $X(b)$ takes values over the whole real line, defining an integration measure for mono-fluids, and using the normalisability condition $|\langle\psi|\psi\rangle| = 1$. The inner product is written in 2 equivalent forms using the general solution for monofluids:
\begin{equation}
    \psi(b, T) = \int \frac{d\alpha}{\sqrt{2\pi\textfrak{h}}}\mathcal{A}(\alpha)\exp\left[\frac{\mathrm{i}}{\textfrak{h}}\alpha(X(b)-T)\right]
\end{equation}
and (\ref{factstate}) is written in equivalent forms \cite{gielen}:
\begin{equation}
    \langle\psi_1|\psi_2\rangle = \int dT\psi_1^*(b,T)\psi_2(b,T)
\end{equation}
\begin{equation}
    \langle\psi_1|\psi_2\rangle = \int d\alpha\mathcal{A}_1^*(\alpha)\mathcal{A}_2(\alpha).
\end{equation}
\newline
\textbf{2) Multifluids - no bounce}\\
As discussed in \cite{gielen}, it is possible to propose that the inner product is given in the form
\begin{equation}
    \langle\psi_1|\psi_2\rangle = \int d\bm{\alpha}\mathcal{A}_1^*(\bm{\alpha})\mathcal{A}_2(\bm{\alpha})
\end{equation}
and since it is time independent, unitarity is preserved. If the wave function takes the form
\begin{equation}
    \psi(b,\bm{T})=\int d\bm{\alpha}\mathcal{A}(\bm{\alpha})\exp\left[ -\frac{\mathrm{i}}{\textfrak{h}}\bm{\alpha}\cdot\bm{T}\right]\psi_s(b;\bm{\alpha})
\end{equation}
Then the inner product will therefore take the form
\begin{equation}
    \langle\psi_1|\psi_2\rangle = \int d\bm{T}d\bm{T}'\psi_1^*(b,\bm{T})\psi_2(b,\bm{T}')K(b,\bm{T}-\bm{T}')
\end{equation}
where
\begin{equation}
    K(b,\bm{T}-\bm{T}')=\int d\bm{\alpha}\frac{e^{-\frac{\mathrm{i}}{\textfrak{h}}\bm{\alpha}\cdot(\bm{T}-\bm{T}')}}{(2\pi\textfrak{h})^{2D}|\psi_s(b;\bm{\alpha})|^2}
\end{equation}
If $\psi_s$ is a pure phase it would not be possible to recover a form like (\ref{firstorwav}). Because equation (\ref{firstorwav}) requires that $\psi_s$ is a plane wave in $X$ and dependent on $b$. The fact that the Kernel is not diagonal for the $X$ inner product induces the ringing quantum effect where the incident and reflected waves interfere.\\
\newline
\textbf{Semiclassical Measure}\\
In \cite{gielen} the wave packet approximation is used, noting that it may erase some quantum information, to make the $b$ calculation straightforward. The bounce is ignored and double branch set up, and the multi-fluids minisuperspace is seen as a dispersive medium that has the single dispersion relation \cite{J2110}
\begin{equation}
    \bm{\alpha \cdot T}-P(b,\bm{\alpha})=0
\end{equation}
Provided the amplitude $\mathcal{A}(\bm{\alpha})$ can be factorised and peaks around $\bm{\alpha}_0$ then $P$ can be Taylor expanded around $\bm{\alpha}_0$ to get
\begin{equation}
    \psi \approx e^{\frac{\mathrm{i}}{\textfrak{h}}(P(b;\bm{\alpha}_0)-\bm{\alpha}_0 \bm{\cdot T})}\prod_i\psi_i(b,T_i)
\end{equation}
With further calculations the probability factorises to 
\begin{equation}
    \mathcal{P}(b,\bm{T})=\prod_{i=1}^D\mathcal{P}_i(b,\bm{T}_i)
\end{equation}
This is normalised using the measure $d\mu_i(b)=dX_i^{eff}$, which implies the probability can be seen as a probability distribution for $b$ at a specific value of $T_i$, that has unspecified values at other times.\\
\newline
\textbf{3) When there's bounce}\\
In the case where $D = 2$ the wave function is dependent on $m$ and $\phi$ \cite{gielen}. Each factor has a superposition of the incident wave (-), the reflected wave (+), and the evanescent wave. The inner product in this approximation is
\begin{equation}
    \langle \psi_1|\psi_2\rangle = \prod_{i=1}^D \left(\int dX^{eff}_{i+}\psi_{i+1}^*(b, T_i)\psi_{i+2}(b, T_i) + \int dX^{eff}_{i-}\psi_{i-1}^*(b, T_i)\psi_{i-2}(b, T_i)\right)
\end{equation}
and the interference between the + and - waves disappears.\\
For $b\geq b_0$ the probability takes the form
\begin{equation}\label{bounceprob}
    \mathcal{P}_i(b;\bm{T}_i)=|\psi_{i+}|^2 \bigg|\frac{dX^{eff}_{i+}}{db}\bigg|+|\psi_{i-}|^2 \bigg|\frac{dX^{eff}_{i-}}{db}\bigg|.
\end{equation}
\begin{equation}
    \frac{dX^{eff}_{1\pm}}{db} = \frac{b^2}{2}\pm\frac{b^4-2m/\phi}{2\sqrt{b^4-b^4_0}}
\end{equation}
The measure factors are:
\begin{equation}
    \frac{dX^{eff}_{2\pm}}{db} = \frac{\mp 1}{\sqrt{b^4-b^4_0}}
\end{equation}
Figure \ref{fig:phenom} shows how the ringing is no longer present in the probability.
\subsection{Towards Phenomenology}
\begin{figure}[ht]
    \centering
    \includegraphics[width=9cm]{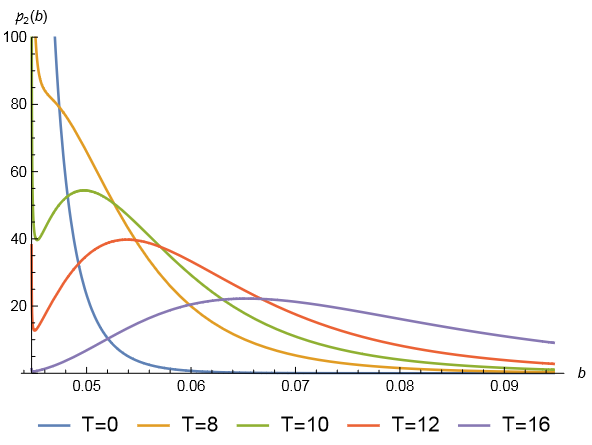}
    \caption{The probability distribution $|\psi|^2$ in the classically allowed region for $b \geq b_0$ at different times (the particular case of $T_m$). Reference: \cite[Figure~9]{gielen}.}
    \label{fig:phenom}
\end{figure}
The figure shows the probability with the semiclassical measure, for different values of $T_m$, using Gaussian wave packets. It is the measure factor that leads to the double peaked distributions. It can be seen that the main peaks are present at $T_m=$ 10, 12, 16.\\
Recalling the classical trajectory is reproduced by $T_m = X^{eff}_{i\pm}$ \cite{J2110}, we can see that as we approach the bounce, the main peak disappears. This is demonstrated in Figure \ref{fig:phenom} when T=8; where the only remaining peak is at ($b=b_0$). The peak becomes more narrow and sharp as $|T_m|\rightarrow0$ so the average value of $b$ becomes greater and greater, becoming larger than the classical trajectory. However the peak will be stuck at $b=b_0$.\\
This leads to the following phenomenology:
\begin{itemize}
    \item $1/b_0$ is the largest classical comoving Hubble length a universe undergoing a $b$-bounce will experience.
    \item The $1/b$ turning point is biased towards this maximum
    \item The strength and length of the effect depends on $\sigma_T$ for the clock being used, which then depends on the sharpness of the conjugate constant.
    \item The sharper the constant, the larger $\sigma_T$, therefore the greater the effect around the $b$-bounce.
\end{itemize}
\newpage
\section{The Evanescent Wave}
\begin{figure}[h]
    \centering
    \includegraphics[width=9cm]{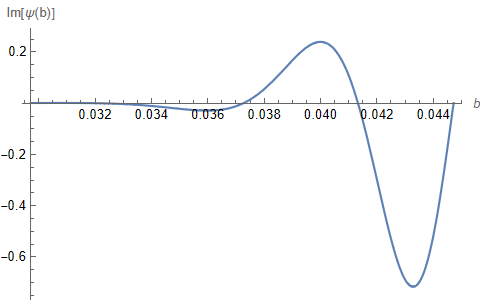}
    \caption{A plot of the imaginary part of the evanescent wave function $\psi_{s-}$, where $b<b_0$, $b_0 = 0.0447$, $m=1$, $T=0$, $\phi = 10^6$, $\textfrak{h} = 1$, and $\sigma_T=4$.}
    \label{fig:psi}
\end{figure}

\begin{figure}[ht]
    \centering
    \begin{subfigure}[b]{7.5cm}
        \centering
        \includegraphics[width=7.5cm]{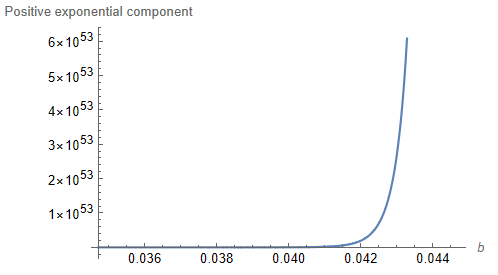}
        \caption{Positive exponential component of $|\psi|^2$}
        \label{fig:pos}
    \end{subfigure}
    \begin{subfigure}[b]{7.5cm}
        \centering
        \includegraphics[width=7.5cm]{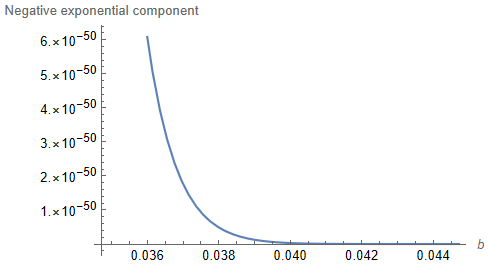}
        \caption{Negative exponential component of $|\psi|^2$}
    \label{fig:neg}
    \end{subfigure}
    \caption{The separated positive and negative components of $|\psi|^2$. It can be seen in Figure \ref{fig:neg} that the exponential decays faster than the exponential diverges in Figure \ref{fig:pos}.}
    \label{fig:exps}
\end{figure}

We've now reached the final part of the thesis where we approach the original concept of the universe quantum tunneling to a different classical region at the time of the $b$-bounce.  From Section \ref{bbounce} we know that when $b<0$ the universe is contracting; if the universe tunneled to $b = -b_0$ during the bounce, it would have tunneled from an expanding state to a contracting state. We first derive the wave function $\psi$ for the evanescent wave packet, and then find probability density $|\psi|^2$ at $-b_0$.
The solution for the evanescent wave function is given in \cite{gielen},
\begin{equation}\label{psiev}
    \psi_{ev}(b) = B\psi_{s-}(b)
\end{equation}
We begin by taking only the $\psi_{s-}$ part of equation (\ref{psis10})
\begin{equation}
    \psi_{s-}(b,\alpha) = \mathcal{N}\exp\left[\frac{\mathrm{i}}{\textfrak{h}}P_-(b,\alpha)\right]
\end{equation}
Using the same factorisable state as in Section \ref{10.4},
\begin{equation}\label{A}
    \mathcal{A}(\alpha) = \prod_i \mathcal{A}_i(\alpha_i)= \prod_i \frac{1}{(2\pi \sigma^2_{\alpha i})^{\frac{1}{4}}} \exp\left[-\frac{(\alpha_i-\alpha_{i0})^2}{4\sigma^2_{\alpha i}}\right]
\end{equation}
we substitute it into (\ref{psiev}) to construct the wave packet, so our wave function now becomes
\begin{equation}\label{psisub}
    \psi_{ev}(b,T) = \int d\alpha \mathcal{A}(\alpha)\exp\left[-\frac{\mathrm{i}}{\textfrak{h}}\alpha\cdot T\right]\exp\left[\frac{\mathrm{i}}{\textfrak{h}}P_-(b,\alpha)\right]
\end{equation}
By Taylor expanding $P_-(b,\alpha)$ around $\alpha_0$, we can approximate
\begin{equation}\label{P-}
    P_-(b,\alpha) \approx P_{-}(b;\alpha_0) + \frac{\partial P_-}{\partial \alpha_i}\bigg|_{\alpha_0}(\alpha - \alpha_0)
\end{equation}
The group speed of the packets is set to one by using the linearising variable \cite{gielen}
\begin{equation}\label{X-}
    X^{\mbox{eff}}_- = \sum_i \frac{\partial P_-}{\partial \alpha_i}\bigg|_{\alpha_0}
\end{equation}
Next we substitute (\ref{A}) and (\ref{P-}) into (\ref{psisub})
\begin{equation}
 \psi_{ev}(b,T) = a\int^{+\infty}_{-\infty} d\alpha \exp\left[-\frac{(\alpha-\alpha_{0})^2}{4\sigma^2_{\alpha}}\right]\exp\left[-\frac{\mathrm{i}}{\textfrak{h}}\alpha\cdot T\right]\exp\left[\frac{\mathrm{i}}{\textfrak{h}}\left(P_{-0} + X^{\mbox{eff}}_-(\alpha - \alpha_0)\right)\right]
\end{equation}
where $a=\frac{1}{(2\pi \sigma^2_{\alpha})^{\frac{1}{4}}}$, and $P_-(b;\alpha_0)$ as $P_{-0}$. We now transform the coordinates letting $\alpha \rightarrow \alpha - \alpha_0$. Denoting $\alpha - \alpha_0$ as $\Delta\alpha$.
\begin{equation}
\begin{split}
    \psi_{ev}(b,T) & = a\int^{+\infty}_{-\infty} d\Delta\alpha \exp\left[-\frac{\Delta\alpha^2}{4\sigma^2_{\alpha}}\right] \exp\left[-\frac{\mathrm{i}}{\textfrak{h}}\left(\Delta\alpha\cdot T+\alpha_0\cdot T\right)\right]\exp\left[\frac{\mathrm{i}}{\textfrak{h}}\left(P_{-0} + X^{\mbox{eff}}_-\Delta\alpha\right)\right]\\
        & = a\exp\left[\frac{\mathrm{i}}{\textfrak{h}}(P_{-0}-\alpha_0\cdot T)\right]\int^{+\infty}_{-\infty} d\Delta\alpha \exp\left[-\frac{\Delta\alpha^2}{4\sigma^2_{\alpha}}\right]\exp\left[-\frac{\mathrm{i}}{\textfrak{h}}\left(\Delta\alpha\cdot T\right)\right]\exp\left[\frac{\mathrm{i}}{\textfrak{h}}X^{\mbox{eff}}_-\Delta\alpha\right]\\
        & = a\exp\left[\frac{\mathrm{i}}{\textfrak{h}}(P_{-0}-\alpha_0\cdot T)\right]\int^{+\infty}_{-\infty} d\Delta\alpha \exp\left[-\frac{\Delta\alpha^2}{4\sigma^2_\alpha}+\frac{\mathrm{i}}{\textfrak{h}}\left(X^{\mbox{eff}}_--T\right)\Delta\alpha\right]
\end{split}
\end{equation}
We complete the square in the exponential so the wave function becomes
\begin{equation}
\begin{split}
    \psi_{ev}(b,T) & = a\exp\left[\frac{\mathrm{i}}{\textfrak{h}}(P_{-0}-\alpha_0\cdot T)\right]\int^{+\infty}_{-\infty} d\Delta\alpha \exp\left[-\frac{1}{4\sigma^2}\left(\Delta\alpha-\frac{2\sigma^2\mathrm{i}}{\textfrak{h}}\left(X^{\mbox{eff}}_--T\right)\right)^2\right]\exp\left[-\frac{\sigma^2}{\textfrak{h}^2}\left(X^{\mbox{eff}}_--T\right)^2\right]\\
        & = (8\pi\sigma^2)^\frac{1}{4}\exp\left[\frac{\mathrm{i}}{\textfrak{h}}(P_{-}(b; \alpha_0)-\alpha_0T)-\frac{\sigma^2}{\textfrak{h}^2}(T-X^{\mbox{eff}}_-)^2\right]
\end{split}
\end{equation}
Using the Heisenberg relation $\sigma_\phi \sigma_T = \frac{\textfrak{h}}{2}$, we re-write the wave function as
\begin{equation}\label{psisol}
    \psi_{ev}(b,T) = \left(\frac{2\pi \textfrak{h}^2}{\sigma_T^2}\right)^{\frac{1}{4}} \exp\left[\frac{\mathrm{i}}{\textfrak{h}}(P_{-}(b, \alpha_0)-\alpha_0T) - \frac{1}{4\sigma_T^2}(T-X^{\mbox{eff}}_-)^2\right]
\end{equation}
$P_-(b;\alpha_0)$ is given by, 
\begin{equation}
\begin{split}
    P_{-0} & = \phi \int^{b}_{b_0} d \tilde b \frac{V-\mathrm{i}\sqrt{V^2_0-V^2}}{2}
\end{split}
\end{equation}
\textbf{Note:} The limits of integration are written in this order to ensure a negative sign for the real exponential. From now on, $\alpha_0 \equiv m$.\\
From equation (\ref{X-}) we can differentiate $P_{-0}$ with respect to m, since $V^2_0 = \frac{4m}{\phi}$, to get $X_-$
\begin{equation}
\begin{split}
    X^{\mbox{eff}}_{-m} & = \frac{\partial P_-}{\partial m} = \frac{\partial}{\partial m}\left(\phi \int^{b}_{b_0} d \tilde b \frac{V-\mathrm{i}\sqrt{V^2_0-V^2}}{2}\right)\\
        & = \frac{-\mathrm{i}\phi}{2}\int^{b}_{b_0} \frac{4/\phi}{2\sqrt{V^2_0-V^2}}d \tilde b-\frac{\phi}{2}\frac{\partial b_0}{\partial m}V^2_0\\
        & = - \mathrm{i}\int^{b}_{b_0}\frac{1}{\sqrt{V_0^2-V^2}}d\tilde b-\frac{\phi}{2}\frac{\partial b_0}{\partial m}V^2_0
\end{split}
\end{equation}
The second term comes from differentiating the $b_0$ limit; since it is independent of b, it can be ignored.\\
\newline
We choose $k = 0$ because the other cases become more complicated as there could be a curvature dominated epoch before the $\Lambda$ domination. Since $k=0$ then $V = b^2$ so now we right $X^{\mbox{eff}}_{-m}$ as
\begin{equation}\label{x-m}
    X^{\mbox{eff}}_{-m} = -\mathrm{i} \int^{b}_{b_0}\frac{1}{\sqrt{b_0^4-\tilde b^4}}d\tilde b.
\end{equation}
The solution for the evanescent wave in equation (\ref{psisol}) is plotted here in Figure \ref{fig:psi}; we use the value of $b_0 = (4 m / \phi)^{1/4}=0.0447$ from \cite{gielen}.
\newpage
Computing $|\psi|^2$ we get
\begin{figure}[h]
    \centering
    \includegraphics[width=9cm]{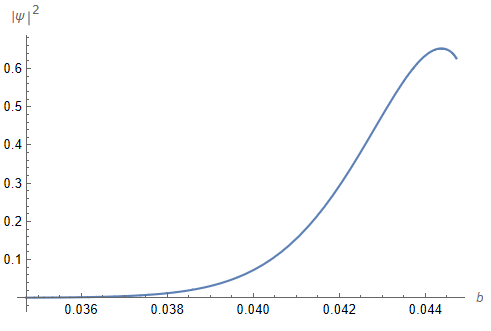}
    \caption{The probability distribution of $|\psi|^2$ between $0.0347$ and $0.0447$ ($(b_0-0.01) \rightarrow b_0$).}
    \label{fig:prob}
\end{figure}
\newline
\begin{figure}[h]
    \centering
    \includegraphics[width=9cm]{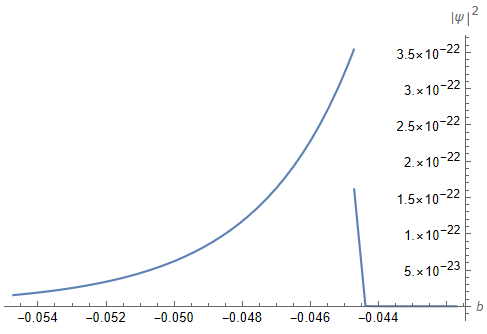}
    \caption{The probability distribution of $|\psi|^2$ between $-0.0547$ and $-0.0417$ ($(-b0 - 0.01) \rightarrow (-b0 + 0.003$)).}
    \label{fig:piecewise}
\end{figure}
\newline
We can see that the Gaussian in our wave function diverges with the $\exp[X^2_-]$ component, however from Figure \ref{fig:prob}, it is clear that $|\psi|^2$ doesn't diverge. This must be because the plane wave decays faster and so cancels out the divergence. This can be done analytically by separating the exponential into positive and negative components as demonstrated in Figure \ref{fig:exps}. Figure \ref{fig:piecewise} reveals the piecewise nature of the probability distribution at $-b_0$, this is expected since $\frac{\partial |\psi|^2}{\partial b}$ is undefined at $-b_0$.\\
\newline
It is not possible to use equation (\ref{bounceprob}) to calculate the probability, as this is only valid for $b\geq b_0$. So we compute the probability density $|\psi|^2$ at $-b_0$ which is $|\psi|^2_{-b_0} = 3.54306 \times 10^{-22}$. This gives an indication of the probability of the universe tunneling from one classical region to the other; finding the probability is left for future works.\\
\newline
\begin{figure}
    \centering
    \includegraphics[width = 10cm]{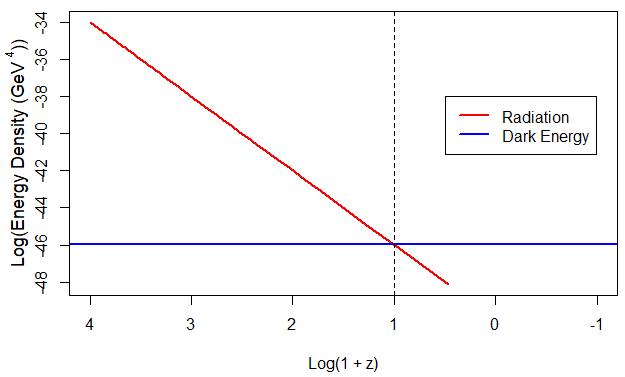}
    \caption{The evolution of radiation, and dark energy densities with redshift. The dotted line shows the location of the bounce, where the incident and reflected wavelengths interfere.}
    \label{fig:ed}
\end{figure}
\newpage
Figure \ref{fig:ed} shows the evolution of the energy densities of radiation and dark energy. When the universe tunnels, instead of going from the expanding universe that was in the radiation epoch on the path following the red line towards the bounce, where the dark energy, shown by the blue line, then would have become the dominant factor. It tunnels to the contracting universe $b < 0$, where it was in the dark energy epoch; following the blue line towards he bounce and then going into the radiation epoch.
\newpage

\section{Conclusion}
In this dissertation we began by covering how the wave function of the universe is obtained either by solving the Wheeler-DeWitt equation or by evaluating a path integral. The bubble universe nucleation was used as an analogy for the tunneling wave function; this analogy provided important insights, such as opening a realm to learn about 2-dimensional quantum gravity. The observer in the analogy exists in the target space, whereas in quantum cosmology the observer exists on the worldsheet. Relating the wave functions of the different observers is an area for future research.\\
The equivalence of the Hartle-Hawking and tunneling wave functions condemns the "creation of the universe from nothing" ($a=0$), posing the question as to whether the Euclidean section ($a^2 < 0$) should be included. Section 9 covered the discussion of how the transition of states is seen as a bounce in connection space. In Section 10 we covered work on the quantum effects of the bounce in $b$, and demonstrated how to construct the wave packets which we used in Section 11. 
The main goal of this thesis was to investigate how the universe could tunnel from one classical region to another at the $b$-bounce. To the best of our knowledge this is the first time that the universe tunneling at a connection bounce has been explored. Although this study is limited by the use of a toy model, it is still useful because it provides a solid foundation for future research; a natural progression of this work would be to use a more realistic model. We found the probability density at $-b_0$ to be $|\psi|^2_{-b_0} = 3.54306 \times 10^{-22}$; this indicates the probability that at the moment of the bounce, it tunnels to another classical region where the universe is contracting and transitioning from $\Lambda$ to radiation.

\newpage
\bibliographystyle{ieeetr} 
\bibliography{refs}

\end{document}